\documentclass[showpacs,preprintnumbers,amsmath,amssymb]{revtex4}

\usepackage{graphicx}
\usepackage{dcolumn}
\usepackage{bm}
\usepackage{color}

\begin{document}

\title{Exclusion of $c\bar c$ Interpretation for $X(3940)$}
\author{W. Sreethawong, K. Xu, Y. Yan
\vspace*{0.4\baselineskip}}
\affiliation{
School of Physics, Suranaree University of Technology, \\
111 University Avenue, Nakhon Ratchasima 30000, Thailand \\
\\
\vspace*{0.4\baselineskip}}

\begin{abstract}
Partial decay widths of the $X(3940)$ are evaluated in the
$^3P_0$ quark model, assuming a charmonium scenario for its structure.
In the study all model parameters are predetermined
by other reactions.
The work reveals that it is difficult to accommodate the $X(3940)$
with any $c\bar c$ meson state in the picture of the potential quark model
plus the $^3P_0$ quark dynamics.
\end{abstract}

\maketitle

In the past decade, many new charmonium-like and bottomonium-like states referred to $XYZ$ have been observed experimentally.
Their production processes include the double charmonium production, the B meson decays, the two photon fusion, the initial state radiation and the decays of excited charmonium-like/bottomonium-like states. A review and updated lists of $XYZ$ states can be found in \cite{rev_Chen, rev_Liu,xyzlist}. Many of these new states are unexpected states, which do not fit into a simple picture of quark model \cite{gellmann}. In this respect, theoretical and experimental investigations of the pattern of $XYZ$ spectra are expected to provide a gateway to understanding the nonperturbative behavior of QCD.

The $X(3872)$ is the first observed charmonium-like state, reported by the Belle Collaboration in 2003 \cite{x3872Belle1}, and has been extensively investigated. The state $X(3872)$ was observed in the $\pi^+\pi^-J/\psi$ mode and later confirmed by the Belle \cite{x3872Belle2}, CDF \cite{x3872cdf}, D0 \cite{x3872d0}, BABAR \cite{x3872babar}, LHCb \cite{x3872lhcb}, and CMS \cite{x3872cms} collaborations in more decay channels.
It has been pointed out in many works that $X(3872)$ is not consistent with a simple $c\bar c$ picture. The observed mass of $X(3872)$ differs from the mass of charmonium in the quark model calculation by $\sim\mathcal{O}(100$ MeV). In addition, the comparable rates of
the $X\to\rho J/\psi$ and $X\to\omega J/\psi$ decays imply strong violation of isospin symmetry, so the $X(3872)$ is identified as an exotic state. Several theoretical explanations of the structure of $X(3872)$ have been proposed including the molecular state \cite{Close,Voloshin,Wong,Swanson3872,Tornqvist}, the tetraquark state \cite{Maiani, Matheus,tetraquark}, and the hybrid charmonium \cite{hybrid}. Among these proposals, the loosely bound $D\bar D^*$ molecular description is the most popular one since the $X(3872)$ lies extremely close to the $D\bar D^*$ threshold. Different interactions between $D$ and $\bar D^*$, including the single pion exchange \cite{single_pion}, both pion and quark exchange \cite{Swanson3872}, the pion and sigma meson exchange \cite{pion_sigma}, and the general pseudoscalar, scalar and vector meson exchanges \cite{Lee3872}, have been investigated.

The molecular picture naturally explains both the proximity
of $X(3872)$ to the $D^0\bar D^{*0}$ threshold and the isospin violating
$J/\psi\rho$ decay mode, but fails sharply to understand the experimental data of branching ratios $BR[X(3872)\rightarrow\gamma J/\psi]/BR[X(3872)\rightarrow\pi^+\pi^- J/\psi]$ and $BR[B^0\rightarrow X(3872)K^0]/BR[B^+\rightarrow X(3872)K^+]$.
Instead, it is proposed in \cite{Suzuki,Meng3872,Kalashnikova} that the $X(3872)$ may have a dominant $c\bar c$ component with some admixture of $D^0\bar D^{*0}$.
The detailed analysis of the structure of $X(3872)$ as a composite state containing both hadronic molecular ($DD^*, J/\psi\rho, J\psi\omega$) and $c\bar c$ components has been studied in \cite{mixed_molecular}.
The strong and radiative decays $X(3872)\to J/\psi\gamma,\psi(2S)\gamma,J/\psi\pi^+\pi^-,J/\psi\pi^+\pi^-\pi^0,J/\psi\pi^0\gamma$ have been calculated. It has been shown that recent experimental data support a mixture of charmonium and molecular interpretation of the $X(3872)$.

For tetraquark interpretation, the $X(3872)$ is predicted to be a $J^{PC}=1^{++}$ tetraquark state with the symmetric spin distribution \cite{Maiani}. However, this proposal was excluded later by the BABAR collaboration due to the negative result of such charged partners \cite{charged_x3872babar}. Following Ref. \cite{Maiani}, the mass of $X(3872)$ is studied with QCD spectral sum rules by including the contributions of higher dimension condensates \cite{Matheus}. Thereafter, the QCD sum rules have been extensively utilized to study the hidden charmed/bottom tetraquark states.

Another very interesting exotic state is $Z(4430)$. It was firstly reported by the Belle Collaboration in $B\to K\pi^+\psi'$ \cite{z4430belle}.
Even though it was not confirmed by BaBar \cite{z4430babar}, the recent observation by LHCb supported the existence of the state \cite{z4430lhcb}. It is a charged state and therefore cannot be described as a conventional charmonium or charmonium-like state. Several theoretical investigations have been carried out to explain and predict the properties and structure of $Z(4430)$. One suggestion is that $Z(4430)$ could be a genuine tetraquark state with $[cu][\bar c\bar d\,]$ diquark anti-diquark content \cite{Maiani2007}. Tetraquark interpretation is also proposed based on QCD-string model \cite{z4430string}. Additionally, $D^*(2010)\bar D_1(2420)$ molecular picture has also been proposed \cite{Rosner,Meng4430,Branz,Wang}. Other theoretical interpretations include cusp effect \cite{Bugg}, $\Lambda_c\Sigma^0_c$ baryonium state \cite{Qiao}, $D_s$ radial excitation \cite{Matsuki}, as well as QCD sum rule studies based on $D^*\bar D_1$ molecule \cite{Lee, Braaten, Bracco}.

In this work, we are interested in the $X(3940)$ state. It was first observed by the Belle Collaboration as an enhancement
at ($3943\pm 6\pm 6$) MeV in the spectrum of mass recoiling against the $J/\psi$
 in the process $e^+e^-\to J/\psi\,X$ via $X(3940)\to D^*\bar D$ decay mode \cite{Belle2007}.
The decay width of the state is determined to be less than 52 MeV at the 90\% C.L.
A new measurement of the $X(3940)$ was performed in the processes $e^+e^-\to J/\psi\,D^{(*)}\bar D^{(*)}$
later by the same
collaboration \cite{Belle2008} and the mass and width were reported $M=(3942^{+7}_{-6}\pm6)$ MeV and
$\Gamma=(37^{+26}_{-15}\pm8)$ MeV. It is noted that the $X(3940)$ has been observed in the $D^*\bar D$ channel
but neither in the $D\bar D$
nor the $\omega J/\psi$ decay mode. Since all lower-mass states observed in the process $e^+e^-\to J/\psi\,X$ recoiling against $J/\psi$
have $J=0$ (the
$\eta_c(1S)$, $\chi_{c0}$ and $\eta'_c(2S)$ as shown in Fig. 1 of Ref. \cite{Belle2007}), it is natural to
propose the $X(3940)$ to be $\eta''_c(3S)$ \cite{Rosner2005}.
The reaction $e^+e^-\to J/\psi\,X(3940)$ was studied in the framework of light cone formalism \cite{Braguta2006},
supposing that the $X(3940)$ is a $3^1S_0$ state or one of the $2^3P$ states. The results suggested
 that the $X(3940)$ is a $3^1S_0$ charmonium.

Alternative explanation of the nature of the $X(3940)$ as a hybrid charmonium state was suggested in \cite{Petrov3940}. In molecular charmonium study, it is found that $X(3940)$ can be described as a mixed charmonium-$DD^*$molecular state with $J^{PC}=1^{++}$ \cite{Fernandez}.
For tetraquark description, it was shown in \cite{Stancu3940} that the mass of $X(3940)$ resonance does not fit into the $c\bar cq\bar q$ state as a $J^{PC}=2^{++}$ state. Besides, the tetraquark picture may be excluded as there is no experimental evidence on its charged partner.

In this work, we study the decay reactions $X(3940) \to D^{(*)}\bar D^{(*)}$, assuming a $3S$, $2P$ or $2D$ $\bar cc$ state to the $X(3940)$.
We work in the nonperturbative $^3P_0$ quark dynamics in which a $q\bar q$ pair is created from or destroyed into vacuum. The model,
originally introduced by Micu \cite{Micu}, has made considerable
successes in understanding low-energies hadron physics. At least for meson
decay, this approximation has been given a rigorous basis in
strong-coupling QCD. The $^3P_0$ model was first applied to evaluate
strong decay partial widths of the three $c\bar c$ states $\psi(3770)$,
$\psi(4040)$, and $\psi(4415)$ in 1970's \cite{Yaouanc1977a, Yaouanc1977b}.
Barnes {\it et al.} calculated recently in the $^3P_0$ model all open-charm strong decay
widths of 40 $c\bar c$ states up to 4.4 GeV, where a universal coupling strength is employed for the $^3P_0$ vertex and
all mesons take spherical harmonic oscillator wave functions with a single length parameter \cite{Barnes2005}.	

We intend to evaluate the open-charm partial decay widths of the $X(3940)$ in the $^3P_0$ model with all model
parameters well predetermined. The effective vertex of the $^3P_0$ model takes the form as in Refs. \cite{yanNNbar,Kittimanapun},
\begin{eqnarray} \label{eq::1}
    V_{ij}&=&\lambda\,\vec\sigma_{ij}\cdot(\vec p_i-\vec
    p_j)\hat{F}_{ij}\hat{C}_{ij}\delta(\vec p_i+\vec
    p_j)\nonumber\\
    &=&\lambda\sum_\mu\sqrt\frac{4\pi}{3}(-1)^\mu\sigma_{ij}^\mu Y_{1\mu}
    (\vec p_i-\vec p_j)\hat F_{ij}\hat C_{ij}\delta(\vec p_i+\vec
    p_j)
 \end{eqnarray}
where $\sigma^\mu_{ij},\hat{F}_{ij},\hat{C}_{ij}$ are respectively the spin, flavor and color operators
projecting a $q\bar q$ pair to the respective vacuum quantum numbers.
The wave functions of all mesons are approximated with the Gaussian form,
\begin{eqnarray} \label{eq::2}
\Psi_{nlm}(\vec p)=N_{nl}e^{-b^2p^2/2}\,L^{l+1/2}_{n}(b\,p)\,Y_{lm}(\theta,\phi)
\end{eqnarray}
where $L^{l+1/2}_{n}(x)$ are the generalized Laguerre polynomial, $\vec p$ is the
relative momentum between the quark and antiquark in a meson, and $b$ is the
length parameter of the Guassian-type wave function.

There are three model parameters, the length parameters of the radial wave functions of
the $D^{(*)}$ and $X(3940)$ mesons and the effective coupling strength $\lambda$
of the $^3P_0$ vertex, which must be nailed down before the $X(3940)$ is studied.

The $D^{(*)}$ length parameter can be determined with the process
$D^+\to\mu^+\nu_\mu$. The partial decay width of the reaction $D^+\to\mu^+\nu_\mu$
is given by
\begin{eqnarray}
\Gamma = \frac{p_f}{32\,M_D\,\pi^2}\int |T_{D^+\to\mu^+\nu_\mu}|^2 d\Omega
\end{eqnarray}
with
\begin{eqnarray}
T_{D^+\to\mu^+\nu_\mu}=\int \frac{d\vec p}{(2\pi)^{3/2}}\psi(\vec p)\frac{\sqrt{2M_D}}{\sqrt{2E_1}\sqrt{2E_2}} T_{c\bar d\to\mu^+\nu_\mu}
\end{eqnarray}
where $T_{c\bar d\to\mu^+\nu_\mu}$ is the transition amplitude of the process $u\bar d\to\mu^+\nu_\mu$ and $\psi(\vec p)$
is the $D$ meson wave function in momentum space. Used as inputs
the weak coupling constant $G=1.166\times 10^{-5}$ GeV$^{-2}$, the CKM element $|V_{cd}|=0.230$, the $D^+$ meson mass $M_D=1.870$ GeV, the
$c$ quark mass $m_c=1.27$ GeV, the $d$ quark mass as the constituent mass $m_d=0.35$ GeV, and the experimental
value of $\Gamma_{D^+\to\mu^+\nu_\mu}=2.42\times 10^{-7}$ eV, we derive the length parameter of
the $D$ meson to be $B_D=2.28$ GeV$^{-1}$. This value
is larger than the one employed in \cite{Barnes2005}, i.e. 2.0 GeV$^{-1}$. Note that it is impossible to estimate an error range for the the length parameter as
the CKM element $|V_{cd}|$ alone would lead to a sizable error bar for the $D$ meson decay width.

The investigation of the reactions $\psi(2S)\to e^+e^-$ and
$\psi(3770)\to e^+e^-$ reveals that the $\psi(2S)$ possess
a small D-wave component but the $X(3770)$ is mainly a $1D$ state
with some S-wave component. These two mesons may couple as
\begin{eqnarray}\label{mixture}
\psi(2S)=\cos\theta\,|2S\rangle-\sin\theta\,|1D\rangle \nonumber\\
\psi(3770)=\sin\theta\,|2S\rangle+\cos\theta\,|1D\rangle
\end{eqnarray}
where $\theta$ is the mixing angle between the $2S$ and $1D$ states.
By fitting the theoretical decay widths to the experimental values \cite{PDG} of $\Gamma_{\psi(2S)\to e^+e^-}=(2.35\pm 0.04)$ keV and $\Gamma_{\psi(3770)\to e^-e^+}=0.262\pm 0.018$ keV, the length parameter and mixing angle were obtained as
$B_{\psi(3770)}=(1.95\pm 0.01)$ GeV$^{-1}$ and $\theta=(10.69\pm 0.63)^o$ or $(-27.6\pm 0.69)^o$ in the present model \cite{Ayut}.
The mixing angle derived here is in agreement with results from other works \cite{Kuang1990,Rosner2001}.

Using the mixing angle $\theta=10.7^o$ or $-27.6^o$, the length parameters $B_{\psi(3770)}=1.95$ GeV$^{-1}$
and $B_{D}=2.28$ GeV$^{-1}$ to fit the experimental decay widths of
 $\Gamma_{\psi(3770)\to D^+D^-}=(11.15\pm 1.09)$ MeV
and $\Gamma_{\psi(3770)\to D^0\bar D^0}=(14.14\pm 1.36)$ MeV \cite{PDG}, we get the corresponding effective coupling
strength $\lambda=(0.68\pm 0.04)$ or $(4.15\pm 0.20)$.
\begin{figure}
\begin{center}
  \centering
  \includegraphics[width=0.5\textwidth]{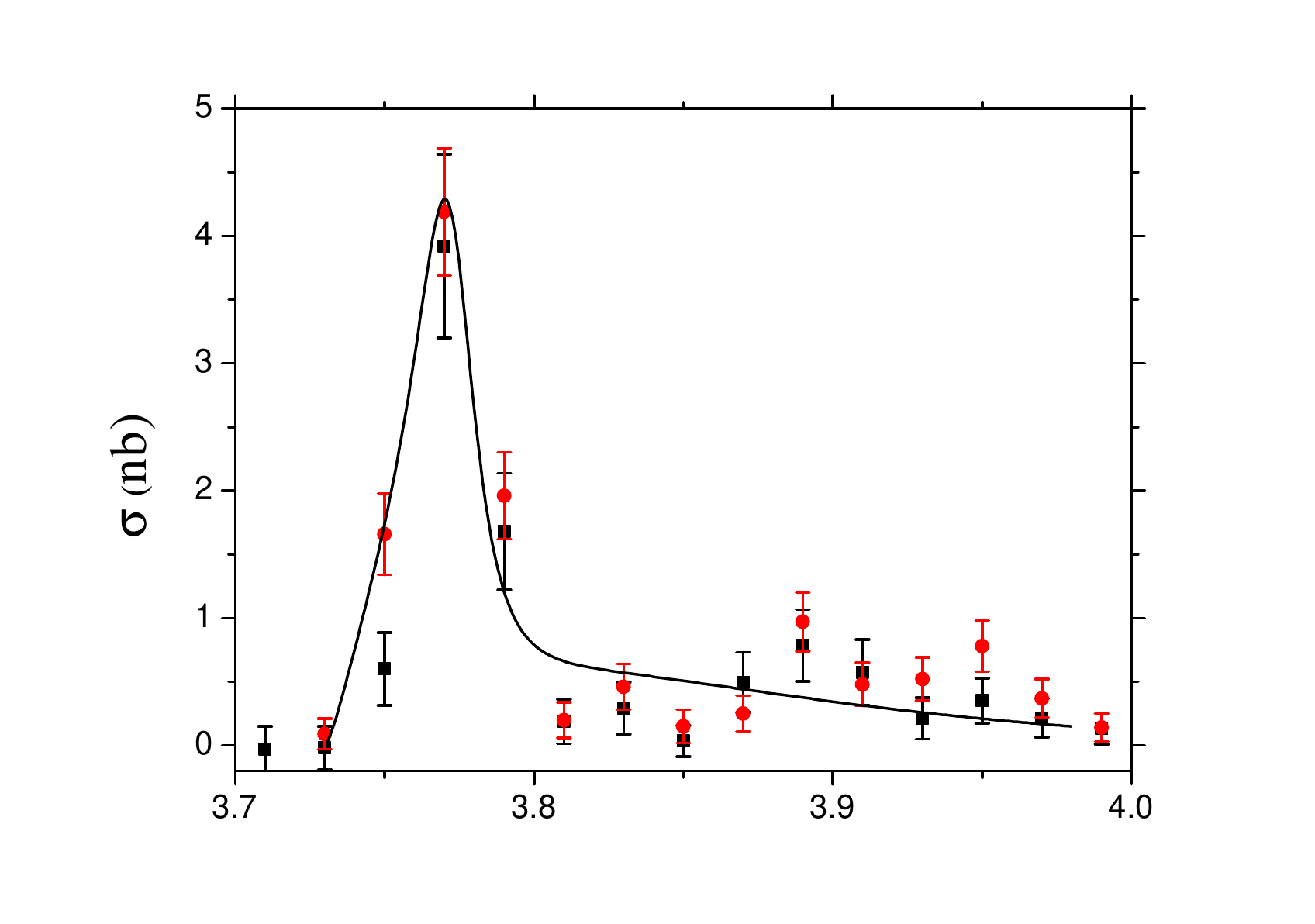} \\
  \vspace{-\baselineskip}
  \includegraphics[width=0.5\textwidth]{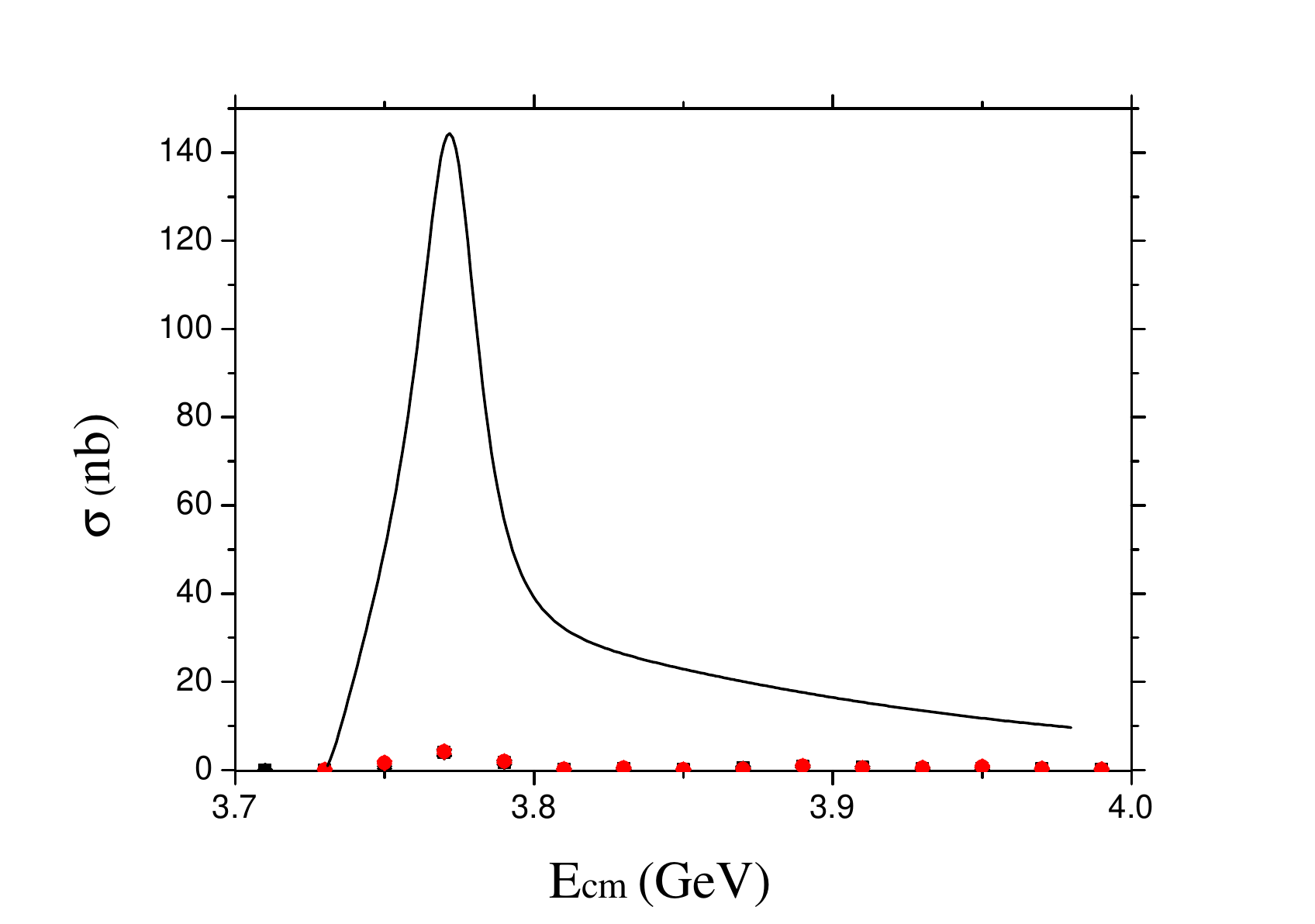}
  \caption{\label{fig1} %
  Theoretical results for the cross section of the reaction $e^+e^-\to D\bar D$ at energy threshold.
  The $J/\psi$, $\psi(2S)$ and $\psi(3770)$ are
included as the intermediate mesons. The upper panel is resulted from the parameters $\{\theta=10.69^o,\,\lambda=0.68,\,B_D=2.28\,{\rm GeV}^{-1},\,B_{\psi(3770)}=1.95\,{\rm GeV}^{-1}\}$ while the lower panel is from the parameters $\{\theta=-27.6^o,\,\lambda=4.15,\,B_D=2.28\,{\rm GeV}^{-1},\,B_{\psi(3770)}=1.95\,{\rm GeV}^{-1}\}$.
    The experimental data are from the Belle \cite{Belle1} and the BaBar \cite{BaBar1}.}
\end{center}
\end{figure}

We expect that the reaction $e^+e^-\to D\bar D$ at energy threshold is dominated by the two-step process in accordance
with other work using an effective Lagrangian approach \cite{Zhang} and with studies of various reactions at low energies in the $^3P_0$ models \cite{yanNNbar,Kittimanapun,yan2010}. In the two-step process, the
$e^+e^-$ pair annihilates into a virtual time-like photon which decays into a $c\bar c$ pair; these
created $c\bar c$ firstly form an intermediate vector meson and then the vector meson finally
decays into $D\bar D$. With only the $J/\psi$, $\psi(2S)$ and $\psi(3770)$
included as intermediate mesons, we fit the lineshape of  the $\psi(3770)$ meson in the $e^+e^-\to D\bar D$ cross section
with two sets of model parameters: $\{\theta=10.69^o,\,\lambda=0.68,\,B_D=2.28\,{\rm GeV}^{-1},\,B_{\psi(3770)}=1.95\,{\rm GeV}^{-1}\}$ and $\{\theta=-27.6^o,\,\lambda=4.15,\,B_D=2.28\,{\rm GeV}^{-1},\,B_{\psi(3770)}=1.95\,{\rm GeV}^{-1}\}$
as shown in Fig. \ref{fig1}.
 It is found that the experimental data strongly
favor the first set of parameters as the second set of parameters leads to a $\psi(3770)$
peak over 10 times higher than the experimental values \cite{Belle1, BaBar1}.

Next, we evaluate the partial decay widths of the $X(3940)$ meson in the $^3P_0$ quark model with all model parameters predetermined. We assume in this work that the length parameter of X(3940) is given by $B_{X(3940)}=B_{\psi(3770)}$=1.95 GeV$^{-1}$ and we use $B_{D^{(*)}}$=2.28 GeV$^{-1}$
and the effective coupling strength of the $^3P_0$ vertex $\lambda=0.68\pm 0.04$.
We consider the $X(3940)$ as either a $3S$, $2P$ or $2D$ $c\bar c$ meson since
potential models predict a mass around 4 GeV for a $c\bar c$ meson in the $3S$, $2P$ and $2D$ states.
Listed in Table I are the theoretical results compared with the Belle data \cite{Belle2008}
in the last row. The values in parentheses are decay widths derived with the parameters in Ref. \cite{Barnes2005}, where the $^3P_0$ strength $\lambda=0.4$ and a size parameter 2.0 GeV$^{-1}$ is applied to all mesons.
It is noted that our results are fairly consistent with the ones derived with the model parameters from
Ref. \cite{Barnes2005}.

\begin{table}
\begin{center}
\label{rhoomega} \caption{Partial decay widths of the processes $X(3940)\to D^{(*)}\bar D^{(*)}$, where $X(3940)$ is assumed to be a $3S$, $2P$ or
$2D$ $c\bar c$ state. The Belle
data \cite{Belle2008} is shown in the last row.}
\vspace*{.3cm}
\begin{tabular}{lllccc}
\hline
\hline
\\
& States && $\Gamma_{D\bar D}$ (MeV) &{}& $\Gamma_{D^*\bar D}$  (MeV)  \\
\\
\hline
\\
& $3\,^1S_0$ && - && $99.8\pm 12.0\, (74.1)$ \\
\\
& $3\,^3S_1$ && $80.3\pm 9.6\, (56.2)$ && $66.5\pm 8.0\, (49.4)$ \\
\\
& $2\,^1P_1$ && - && $109.3\pm 13.1\, (140.4)$  \\
\\
& $2\,^3P_0$ && $90.5\pm 10.9\, (130.8)$ && - \\
\\
& $2\,^3P_1$ && - && $204.5\pm 24.5\, (271.4)$ \\
\\
& $2\,^3P_2$ && $65.7\pm 7.9\, (66.2)$ && $8.5\pm 1.0\, (8.7)$  \\
\\
& $2\,^1D_2$ && - && $15.3\pm 1.8\,(17.2)$  \\
\\
& $2\,^3D_1$ && $54.6\pm 6.6\,(64.3)$ && $12.5\pm 1.5\,(16.1)$  \\
\\
& $2\,^3D_2$ && - && $22.6\pm 2.7\,(25.6)$  \\
\\
& $2\,^3D_3$ && $7.6\pm 0.9 \,(6.4)$ && $0.27\pm 0.03\,(0.23)$  \\
\\\hline\\
& $X(3940)$ && - && $37^{+26}_{-15}\pm8$  \\
\\
\hline
\hline
\end{tabular}
\end{center}
\end{table}

Since the $X(3940)$ has been observed in the $D^*\bar D$ channel
but not in the $D\bar D$ decay mode, the states $3\,^3S_1$, $2\,^3P_0$, $2\,^3P_2$, $2\,^3D_1$ and $2\,^3D_3$ are clearly ruled out and possible states are then $3\,^1S_0$, $2\,^1P_1$, $2\,^3P_1$, $2\,^1D_2$ and $2\,^3D_2$.
However, when considering that the reaction $e^+e^-\to J/\psi\,X$ is dominated by the one photon exchange process, the charge conjugation invariance demands that
the $X(3940)$ must be a $C=1$ state. Therefore, the only possible candidates for the $X(3940)$ are the states $3\,^1S_0$, $2\,^3P_1$ and $2\,^1D_2$ which have positive
charge conjugation.

It is difficult to accommodate the $X(3940)$ with the $2\,^1D_2$ state as the mass of the $2\,^1D_2$ $c\bar c$ meson obtained in the framework of potential
approaches is $4150 - 4210$ MeV \cite{Swanson}. The large decay width of the $2\,^3P_1$ derived in this work makes this state unlikely to accommodate the $X(3940)$.
The prediction for the decay width of the $3\,^1S_0$ state is
more or less consistent with the experimental data, and hence one might consider interpreting the $X(3940)$ as a $3\,^1S_0$ $c\bar c$ state.
However, one problem with this interpretation is that the mass of the $3\,^1S_0$ $c\bar c$ meson is derived as $4040-4060$ MeV in potential approaches \cite{Swanson}.
Given the success of the potential quark model (see Ref. \cite{Swanson,Eichten} for a recent review), to assign the $3\,^1S_0$ $c\bar c$ meson a mass
of 3940 MeV will cause great concern about the non-relativistic $c\bar c$ phenomenology.

One may argue that the prediction of the $^3P_0$ quark model may not be so reliable. We have to admit that if the same model parameters are applied to both
the light and heavy meson sectors, one may expect only qualitative results and sometime even misleading results. For instance, the theoretical
$\rho$ decay width in Ref. \cite{Barnes97} is only half its experimental value, and the partial decay width of the process $\psi(4040)\to D\bar D$ in
Ref. \cite{Barnes2005}
is almost zero which is strongly inconsistent with experimental data \cite{Belle1,BaBar1}.
In this work, however, the model parameters have been fitted with very relevant reactions and hence the theoretical results listed in Table I reflect
the physics to a large extent. We strongly argue that the $^3P_0$ quark model should be applied with the model parameters carefully fixed with very relevant
reactions.

In summary we may conclude that it is difficult to accommodate the $X(3940)$, observed in the
process $e^+e^-\to J/\psi\,X$ via the $X(3940)\to D^*\bar D$ decay mode \cite{Belle2007}, with any $c\bar c$ meson state in the
picture of the potential quark model (for charmonium spectrum evaluation) plus the $^3P_0$ quark dynamics (for
decay width evaluation). Given the great success of the potential quark model and the $^3P_0$ quark dynamics and the careful
predetermination of the model parameters, the conclusion may reflect the physics to a large extent though it is model dependent.

The nature of the $X(3940)$ is still an open question. A tetraquark picture may be excluded
as there is no experimental evidence on a charge partner.
Considering the consistence and inconsistence between experimental data and theoretical predictions for the $X(3872)$
in the molecular picture, one may explore the interpretation for the $X(3940)$ as a mixture of a
$3\,^1S_0$, $2\,^3P_1$ or $2\,^3D_2$ $c\bar c$ component and $DD^*$ molecular state.

\section*{ACKNOWLEDGEMENTS}

This work is supported by Suranaree University of Technology and by the Higher Education Research Promotion and National Research University Project of Thailand, Office of the Higher Education Commission.


\begin{thebibliography}{99}

\bibitem{rev_Chen}
W. Chen, W. -Z. Deng, J. He, N. Li, X. Liu, Z. -G. Luo, Z. -F. Sun and S. -L. Zhe, Pos Hadron2013, 005 (2013).

\bibitem{rev_Liu}
X. Liu, Chin. Sci. Bull. {\bf 59}, 3815 (2014).

\bibitem{xyzlist}
G. T. Bodwin, E. Braaten, E. Eichten, S. L. Olsen, T. K. Pedlar, J. Russ, arXiv:1307.7425.

\bibitem{gellmann}
M. Gell-Mann, Phys. Lett. 8, 214 (1964).

\bibitem{x3872Belle1}
S. K. Choi {\it et al.} [Belle Collaboration], Phys. Rev. Lett. {\bf 91}, 262001 (2003).

\bibitem{x3872Belle2}
K. Abe {\it et al.} [Belle Collaboration], arXiv:hep-ex/0505037;
G. Gokhroo {\it et al.} [Belle Collaboration], Phys. Rev. Lett. {\bf 97}, 162002 (2006);
I. Adachi {\it et al.} [Belle Collaboration], arXiv:0809.1224 [hep-ex];
I. Adachi {\it et al.} [Belle Collaboration], arXiv:0810.0358 [hep-ex];
V. Bhardwaj {\it et al.} [Belle Collaboration], Phys. Rev. Lett. {\bf 107}, 091803 (2011).

\bibitem{x3872cdf}	
D. E. Acosta {\it et al.} [CDF II Collaboration], Phys. Rev. Lett. {\bf 93}, 072001 (2004);
A. Abulencia  {\it et al.} [CDF II Collaboration], Phys. Rev. Lett. {\bf 96}, 102002 (2006);
A. Abulencia  {\it et al.} [CDF II Collaboration], Phys. Rev. Lett. {\bf 98}, 132002 (2007);
T. Aaltonen  {\it et al.} [CDF II Collaboration], Phys. Rev. Lett. {\bf 103}, 152001 (2009).

\bibitem{x3872d0}	
V. M. Abazov {\it et al.} [D0 Collaboration], Phys. Rev. Lett. {\bf 93}, 162002 (2004).

\bibitem{x3872babar}
B. Aubert {\it et al.} [BABAR Collaboration], Phys. Rev. D {\bf 71}, 071103 (2005);
B. Aubert {\it et al.} [BABAR Collaboration],  Phys. Rev. D {\bf 71}, 052001 (2005);
B. Aubert {\it et al.} [BABAR Collaboration],  Phys. Rev. D{\bf 73}, 011101 (2006);
B. Aubert {\it et al.} [BABAR Collaboration],  Phys. Rev. Lett. {\bf 96}, 052002 (2006);
B. Aubert {\it et al.} [BABAR Collaboration],  Phys. Rev. D{\bf 74}, 071101 (2006);
B. Aubert {\it et al.} [BABAR Collaboration],  Phys. Rev. D{\bf 77}, 011102 (2008);
B. Aubert {\it et al.} [BABAR Collaboration],  Phys. Rev. D{\bf 77}, 111101 (2008);
B. Aubert {\it et al.} [BABAR Collaboration],  Phys. Rev. Lett. {\bf 102}, 132001 (2009);
P. del Amo Sanchez {\it et al.} [BABAR Collaboration],  Phys. Rev. D{\bf 82}, 011101 (2010).

\bibitem{x3872lhcb}	
R. Aaij {\it et al.} [LHCb Collaboration], Eur. Phys. J. C {\bf 72}, 1972 (2012);	
R. Aaij {\it et al.} [LHCb Collaboration], Phys. Rev. Lett. {\bf 110}, 222001 (2013).

\bibitem{x3872cms}
S. Chatrchyan {\it et al.} [CMS Collaboration], JHEP {\bf 1304}, 154 (2013).

\bibitem{Close}
F.~E.~Close and P.~R.~Page, Phys.\ Lett.\ B {\bf 578}, 119 (2004).

\bibitem{Voloshin}
M.~B.~Voloshin, Phys.\ Lett.\ B {\bf 579}, 316 (2004).

\bibitem{Wong}
C.~-Y.~Wong, Phys.\ Rev.\ C {\bf 69}, 055202 (2004).

\bibitem{Swanson3872}
E.~S.~Swanson, Phys.\ Lett.\ B {\bf 588}, 189 (2004).

\bibitem{Tornqvist}
N.~A.~Tornqvist, Phys.\ Lett.\ B {\bf 590}, 209 (2004).

\bibitem{Maiani}
L. Maiani, F. Piccinini, A. D. Polosa, V. Rique, Phys.\ Rev.\ D {\bf 71}, 014028 (2005).

\bibitem{Matheus}
R.~D. Matheus, S. Narison, M. Nielsen, J.~M. Richard, Phys. Rev. {\bf D 75}, 014005 (2007);

\bibitem{tetraquark}
H. Hogaasen, J.~M. Richard, P. Sorba, Phys. Rev. {\bf D 73}, 054013 (2006);
H.~Hogaasen, E.~Kou, J.~-M.~Richard and P.~Sorba, arXiv:1309.2049 [hep-ph];
D. Ebert, R.~N. Faustov, V.O. Galkin, Phys. Lett. {\bf B 634}, 214 (2006);
N. Barnea, J. Vijande, A. Valcarce, Phys. Rev. {\bf D 73}, 054004 (2006);
T.~W. Chiu {\it et al.}, Phys. Lett. {\bf B 646}, 95 (2007); Phys. Rev. D 73, 111503 (2006), Erratum-ibid. {\bf D 75}, 019902 (2007);
M.~T.~AlFiky, F.~Gabbiani and A.~A.~Petrov, Phys.\ Lett.\ B {\bf 640}, 238 (2006);
S.~Fleming, M.~Kusunoki, T.~Mehen and U.~van Kolck, Phys.\ Rev.\ D {\bf 76}, 034006 (2007);
E.~Braaten, M.~Lu and J.~Lee, Phys.\ Rev.\ D {\bf 76}, 054010 (2007);
C.~Hanhart, Y.~.S.~Kalashnikova, A.~E.~Kudryavtsev and A.~V.~Nefediev, Phys.\ Rev.\ D {\bf 76}, 034007 (2007);
M.~B.~Voloshin,  Phys.\ Rev.\ D {\bf 76}, 014007 (2007).

\bibitem{hybrid}
F. E. Close {\it et al.}, Phys.\ Rev.\ D {\bf 57}, 5653 (1998);
G. Chiladze, A. F. Falk, A. A. Petrov, Phys.\ Rev.\ D {\bf 58}, 034013 (1998);
F. E. Close, S. Godfrey, Phys.\ Lett. B {\bf 574}, 210 (2003);
B.~A.~Li, Phys.\ Lett.\ B {\bf 605}, 306 (2005).

\bibitem{single_pion}
N.~A.~Tornqvist, Nuovo Cim.\ A {\bf 107}, 2471 (1994);
N.~A.~Tornqvist, Z.\ Phys.\ C {\bf 61}, 525 (1994).

\bibitem{pion_sigma}
Y.~-R.~Liu, X.~Liu, W.~-Z.~Deng and S.~-L.~Zhu, Eur.\ Phys.\ J.\ C {\bf 56}, 63 (2008).

\bibitem{Lee3872}
I.~W.~Lee, A.~Faessler, T.~Gutsche and V.~E.~Lyubovitskij, Phys.\ Rev.\ D {\bf 80}, 094005 (2009).

\bibitem{Suzuki}
M. Suzuki, Phys.\ Rev.\ D {\bf 72}, 114013 (2005).

\bibitem{Meng3872}
C. Meng, Y. -J. Gao, K. -T. Chao, hep-ph/0506222.

\bibitem{Kalashnikova}
Y. S. Kalashnikova, Phys. Rev. D {\bf 72}, 034010 (2005).

\bibitem{mixed_molecular}
Y. B. Dong, A. Faessler, T. Gutsche, V. E. Lyubovitskij, Phys. Rev. D {\bf 77}, 094013 (2008);
Y. B. Dong, A. Faessler, T. Gutsche, S. Kovalenko, V. E. Lyubovitskij, Phys. Rev. D {\bf 79}, 094013 (2009);
Y. B. Dong, A. Faessler, T. Gutsche, V. E. Lyubovitskij, J. Phys. G {\bf 38}, 015001 (2011).

\bibitem{charged_x3872babar}
B. Aubert {\it et al.} [BABAR Collaboration], Phys. Rev. D {\bf 71}, 031501 (2005);

\bibitem{z4430belle}
S. -K. Choi {\it et al.} [Belle Collaboration], Phys. Rev. Lett. {\bf 100}, 142001 (2008).

\bibitem{z4430babar}
B. Aubert {\it et al.} [BABAR Collaboration], Phys. Rev. D {\bf 79}, 112001 (2009).

\bibitem{z4430lhcb}
R. Aaij {\it et al.} [LHCb Collaboration], Phys. Rev. Lett. {\bf 112}, 222002 (2014).

\bibitem{Maiani2007}
L. Maiani, A. D. Polosa, V. Riquer, arXiv:0708.3997 [hep-ph].

\bibitem{z4430string}
S. S. Gershtein, A. K. Likhoded, G. P. Pronko, arXiv: 0709.2058 [hep-ph].

\bibitem{Rosner}
J. L. Rosner, Phys. Rev. D {\bf 76}, 114002 (2007).

\bibitem{Meng4430}
C. Meng and K. -T. Chao, arXiv:0708.4222 [hep-ph]

\bibitem{Branz}
T. Branz, T. Gutsche, V. E. Lyubovitskij, Phys. Rev. D {\bf 82}, 054025 (2010).

\bibitem{Wang}
Q. Wang, C. Hanhart, Q. Zhao, Phys. Rev. Lett. {\bf 111}, 132003 (2013);
Q. Wang, C. Hanhart, Q. Zhao, Phys. Lett. B {\bf 725}, 1-3, 106-110 (2013).

\bibitem{Bugg}
D. V. Bugg, J. Phys. G {\bf 35} 075005 (2008).

\bibitem{Qiao}
C. -F. Qiao, J. Phys. G {\bf 35} 075008 (2008).

\bibitem{Matsuki}
T. Matsuki, T. Morii, K. Sudoh, Phys. Lett. B {\bf 669}, 156 (2008).

\bibitem{Lee}
S. H. Lee, A. Mihara, F. S. Navarra, M. Nielsen, Phys. Lett. B {\bf 661}, 28 (2008);
S. H. Lee, K. Morita, M. Nielsen, Nucl. Phys. A {\bf 815}, 29 (2009).

\bibitem{Braaten}
E. Braaten, M. Lu, Phys. Rev. D {\bf 79}, 051503 (2009).

\bibitem{Bracco}
M. E. Bracco, S. H. Lee, M. Nielsen, R. Rodrigues da Silva, Phys. Lett. B {\bf 671}, 240 (2009).

\bibitem{Belle2007}
K. Abe {\it et al.} (Belle Collaboration), Phys. Rev. Lett. {\bf 98}, 082001 (2007).

\bibitem{Belle2008}
P. Pakhlov {\it et al.} (Belle Collaboration), Phys. Rev. Lett. {\bf 100}, 202001 (2008).

\bibitem{Rosner2005}
J. L. Rosner, arXiv:hep-ph/0508155.

\bibitem{Braguta2006}
V. V. Braguta, A. K. Likhoded, A. V. Luchinsky, Phys. Rev. D {\bf 74}, 094004 (2006).

\bibitem{Petrov3940}
A. A. Petrov, J. Phys. Conf. Ser. {\bf 9}, 83 (2005).

\bibitem {Fernandez}
F. Fernandez, P. G. Ortega, D. R. Entem, AIP Conf. Proc. {\bf 1606}, 168 (2014).

\bibitem{Stancu3940}
Fl. Stancu, arXiv:hep-ph/0607077.

\bibitem{Micu}
L. Micu, Nucl. Phys. {\bf B} 10, 521 (1969).

\bibitem{Yaouanc1977a}
A. Le Yaouanc, L. Oliver, O. Pe`ne, and J.-C. Raynal, Phys. Lett. {\bf B 71}, 397 (1977).

\bibitem{Yaouanc1977b}
A. Le Yaouanc, L. Oliver, O. Pe`ne, and J.-C. Raynal, Phys. Lett. {\bf B} 72, 57 (1977).

\bibitem{Barnes2005}
T. Barnes, S. Godfrey, E. S. Swanson, Phys. Rev. {\bf D} 72, 054026 (2005).

\bibitem{yanNNbar}
Y. Yan, C. Kobdaj, P. Suebka, Y. M. Zheng, Amand Faessler, Th. Gutsche, and V. E. Lyubovitskij,
Phys. Rev. D {\bf 71}, 025204 (2005).

\bibitem{Kittimanapun}
K. Kittimanapun, K. Khosonthongkee, C. Kobdaj, P. Suebka, Y. Yan, Phys. Rev. C {\bf 79}, 025201 (2009).

\bibitem{PDG}
J. Beringer {\it et al.} (Particle Data Group), Phys. Rev. D {\bf 86}, 010001 (2012).

\bibitem{Ayut}
A. Limphirat, W. Sreethawong, K.  Khosonthongkee, Y. Yan, Phys. Rev. D {\bf 89}, 054030 (2014).

\bibitem{Kuang1990}
Y. P. Kuang, T. M. Yan, Phys. Rev. {\bf D} 41, 155 (1990).

\bibitem{Rosner2001}
J. L. Rosner, Phys. Rev. {\bf D} 64, 094002 (2001).

\bibitem{Zhang}
Y. J. Zhang and Q. Zhao, Phy. Rev. D{\bf 81}, 034011 (2010).

\bibitem{yan2010}
Y. Yan, K. Khosonthongkee, C. Kobdaj, P. Suebka, J. Phys. G {\bf 37}, 075007 (2010).

\bibitem{Belle1}
G. Pakhlova {\it et al.} (Belle Collaboration), Phys. Rev. D{\bf 77}, 011103 (2008).

\bibitem{BaBar1}
B. Aubert {\it et al.} (BaBar Collaboration), Phys. Rev. D{\bf 76}, 111105 (2007).

\bibitem{Swanson}
E. S. Swanson, Phys. Rep. {\bf 429}, 243 (2006).

\bibitem{Eichten}
E. Eichten, S. Godfrey, H. Mahlke and J. L. Rosner, Rev. Mod. Phys. {\bf 80}, 1161 (2008).

\bibitem{Barnes97}
T. Barnes, F. E. Close, P. R. Page, and E. S. Swanson, Phys. Rev. D{\bf 55}, 4157 (1997).


\end{thebibliography}
\end{document}